# Structural Trends and Itinerant Magnetism of the New Cage-structured Compound HfMn$_2$Zn$_{20}$

Nusrat Yasmin[1], Md Fahel Bin Noor[1], and Tiglet Besara[1,2]

[1] Department of Physics, Astronomy, and Materials Science, Missouri State University, Springfield, MO 65897, USA

[2] Author to whom any correspondence should be addressed: tigletbesara@missouristate.edu

**Abstract**
A new cage-structured compound – HfMn$_2$Zn$_{20}$ – belonging to the $AB_2C_{20}$ ($A$, $B$ = transition or rare earth metals, and $C$ = Al, Zn, or Cd) family of structures has been synthesized via the self-flux method. The new compound crystallizes in the space group $Fd\bar{3}m$ with lattice parameter $a \approx 14.0543(2)$ Å ($Z$ = 8) and exhibits non-stoichiometry due to Mn/Zn mixing on the Mn-site and an underoccupied Hf-site. The structure refines to Hf$_{0.93}$Mn$_{1.63}$Zn$_{20.37}$ and follows lattice size trends when compared to other Hf$M_2$Zn$_{20}$ ($M$ = Fe, Co, and Ni) structures. The magnetic measurements show that this compound displays a modified Curie-Weiss behavior with a transition temperature around 22 K. The magnetization shows no saturation, a small magnetic moment, and near negligible hysteresis, all signs of itinerant magnetism. The Rhodes-Wohlfarth ratio and the spin fluctuation parameters ratio both confirm the itinerant nature of the magnetism in HfMn$_2$Zn$_{20}$.

**Introduction**
The large family of isostructural ternary intermetallic compounds with formula $AB_2C_{20}$ ($A$, $B$ = transition and/or rare earth metals, and $C$ = Al, Zn, or Cd) crystallize in the CeCr$_2$Al$_{20}$-type structure [1] with space group $Fd\bar{3}m$ and exhibit "cages" formed by the element $C$ which contain the loosely-bound elements $A$ and $B$ [2-5]. Even though over a hundred compounds in this 1-2-20 family have been discovered, new ones are still being found and many of them have been explored for various properties. For example, some of the $AB_2C_{20}$ compounds exhibit superconductivity, mainly among the $C$ = Al compounds [6-12] although it has also been observed in $C$ = Zn compounds [13,14]. The superconductivity has been explored in terms of enhanced $T_C$ due to the rattling of the heavier $A$ and $B$ elements within the cages. The rattling of the $A$ and $B$ elements was also the focus for the most recent surge of interest in the $AB_2C_{20}$ compounds: enhanced low-temperature thermoelectricity [15-17]. This was first explored in Yb-comprising Zn compounds, i.e., Yb$B_2$Zn$_{20}$ where $B$ is a transition metal [15,16] but more recently in site-mixed Yb/Ce/Sm [17].

Many more compounds in the $AB_2C_{20}$ family have naturally been explored for their magnetic properties since the elements $A$ and $B$ are transition metals and/or rare earth metals that exhibit magnetism. The vast majority of them comprise a rare earth element (lanthanide or actinide) as the $A$ element [4,18-38]. Very few have a transition metal as the $A$ element; only with zirconium [39] and hafnium [5]. A few compounds have yttrium as the $A$ element [19,40,41], however, Y is generally considered a rare earth element. The $B$ element, on the other hand, is always a transition metal. The compounds which have manganese as the $B$ element, i.e., the $A$Mn$_2$Zn$_{20}$ compounds, have been shown to exhibit itinerant magnetism [28,39-42]. Interestingly, they have all been reported as non-stoichiometric: while Al or In was added to the $A$Mn$_2$Zn$_{20}$ ($A$ = Y, Ce, Pr, Nd, Sm, Gd, Dy, Er, Yb) resulting in Al/Zn or In/Zn mixing [40,41,43], the ZrMn$_2$Zn$_{20}$ exhibits a mixed Mn/Zn site [39], i.e., Zn mixing with Mn in its site residing within a cage.

Itinerant magnetism arises due to the itinerancy – or movement – of electrons, resulting in a magnetic behavior governed by the collective motion of electrons throughout the lattice instead of localized magnetic moments associated with individual ions. The magnetic moments in such materials come from the spins of the itinerant electrons aligning, therefore, lowering the effective magnetic moment (when compared to that from localized magnetism) since the moments are determined by the collective behavior of the electrons and not exclusively by the individual, localized ions.

In this work, we report on the discovery, single crystal growth, and magnetic properties of a new compound in this family, HfMn$_2$Zn$_{20}$, which, similar to its Zr analog, exhibits Mn/Zn mixing. We show that the structure of this new compound follows the expected lattice size trend when compared to the existing Hf$M_2$Zn$_{20}$ ($M$ = Fe, Co, and Ni) [5] with deviations that can be explained by the Mn/Zn mixing. In addition, our magnetic study confirms the itinerant nature of this compound.

**Methods**
*Synthesis.* Single crystals of HfMn$_2$Zn$_{20}$ were grown via the self-flux method. Elements (>99.9%) were measured in atomic ratio of 1:2:60 Hf:Mn:Zn and loaded in a 2 ml alumina crucible, where the excess zinc acts as flux and provides enough reaction bath despite its boil-off at higher temperatures. A second alumina crucible was filled with quartz wool and put as a cap on the top of the reaction crucible. The two crucibles were then placed in a quartz ampoule. To ensure inert conditions of the reaction, the entire assembly was done inside an argon-filled glovebox. The quartz ampoule was subsequently sealed under vacuum using an oxygen-hydrogen torch and put inside a muffle furnace. The reaction was heated up at 800°C at a rate of 80°C/h, maintained at 800°C for 20 hours, and then slowly cooled to 550°C at a rate of 3°C/h. After completion of the temperature cycle, the quartz ampoule was rapidly taken out of the furnace, flipped upside down, and centrifuged to segregate the crystal from excess molten flux. After centrifuging and cooling, the quartz was broken to extract the crystals from inside the crucible. The extracted crystals had a layer of extra flux on their surface which was etched away with highly dilute HCl solution which attack elemental zinc at a higher rate than the crystals.

*Energy dispersion spectroscopy.* For energy dispersive spectroscopy (EDS)-based elemental analysis, a FEI QUANTA 200 FEG scanning electron microscope (SEM) equipped with an Oxford Instruments Ultim Max EDS detector was used. Several crystals of different sizes and shapes were picked to confirm the stoichiometry of the compound. All the crystals were fixed on copper tape so that a flat surface faced up.

*Single crystal X-ray diffraction.* Single-crystal X-ray diffraction was performed with a Rigaku-Oxford Diffraction XtaLAB Synergy-S diffractometer equipped with a HyPix-6000HE Hybrid Photon Counting detector and dual PhotonJet-S Mo/Cu 50W Microfocus X-ray sources. Data were collected at room temperature with Mo $K\alpha$ radiation ($\lambda$ = 0.71073 Å) using $\omega$ scans with 0.5° frame widths to a resolution of 0.5 Å, equivalent to $2\theta \approx 90°$. Reflections were recorded, indexed, and corrected for absorption using Rigaku Oxford Diffraction CrysAlisPro [44]. The structure refinement was done with CRYSTALS [45], employing the charge-flipping software SUPERFLIP [46] to solve the structure. The data quality permitted an unconstrained full matrix refinement against $F^2$ (the square of the structure factors which are directly proportional to the scattered intensities) with anisotropic displacement parameters for all atoms. A CIF has been deposited with the Cambridge Crystallographic Data Center (CSD #2212294) [47].

*Powder X-ray diffraction.* Powder X-ray diffraction was performed with a Bruker Discover D8 in a Bragg-Brentano geometry, operating at 40 kV and 40 mA. A pattern was collected at room temperature with Cu $K\alpha$ radiation ($\lambda$ = 1.541 Å) in approximately 0.02° steps from 10° to 90° in $2\theta$. The sample comprised several clean crystals that were finely ground and deposited on a glass slide. The pattern was refined with the Rietveld method using the Bruker software TOPAS [48].

*Magnetization.* Magnetic measurements were done in a Quantum Design MPMS-3 system at the Cornell Center for Materials Research. Magnetization as a function of temperature was obtained at an applied field of 0.1 T and the temperature was swept at a rate of approximately 0.1 K/s from 1.8 K to 300 K. Magnetization as a function of field was obtained at temperatures 1.8 K and 300 K and the field was swept at a rate of approximately 50 Oe/s. All measurements were done on a zero field-cooled single crystal of approximately 14.6 mg.

**Result and discussion**

*Structure*

Several crystals of various sizes were grown out of the flux. The crystals exhibit in general an octahedral morphology (double pyramid shape), with the largest crystal approximately 1.5 mm long. Single crystal X-ray diffraction reveals that this compound crystallizes in the cubic space group $Fd\bar{3}m$ ($Z$ = 8), isostructural with all other members of the $AB_2C_{20}$ family [3,5], comprising of one Hf site, one Mn site, and three distinct Zn sites. The unit cell comprises large cages ("voids") of Zn with Hf in the center of a 16-atom coordinated Frank-Kasper polyhedron formed by twelve Zn1 and four Zn3, and Mn in the center of a 12-coordinated icosahedron formed by six Zn1 and six Zn2. All cages are corner-shared via Zn atoms. The Hf-centered Frank-Kasper polyhedra are corner-shared with each other via the Zn3 atoms, the Mn-centered icosahedra are corner-shared with each other via the Zn2 atoms, and the two different polyhedra are corner-shared with each other via the Zn1 atoms. This results in a Hf–Hf distance of 6.086 Å, a Mn–Mn distance of 4.969 Å, and a Hf–Mn distance of 5.827 Å. These structures have also been previously described in detail [3,5,16,39]. Figure 1(a) shows the unit cell along [110] with the Frank-Kasper polyhedra in blue and the icosahedra in purple. Figure 1(b) shows an image of the crystal, obtained with a scanning electron microscope in backscatter mode. Figure 2 displays the two polyhedra along [111].

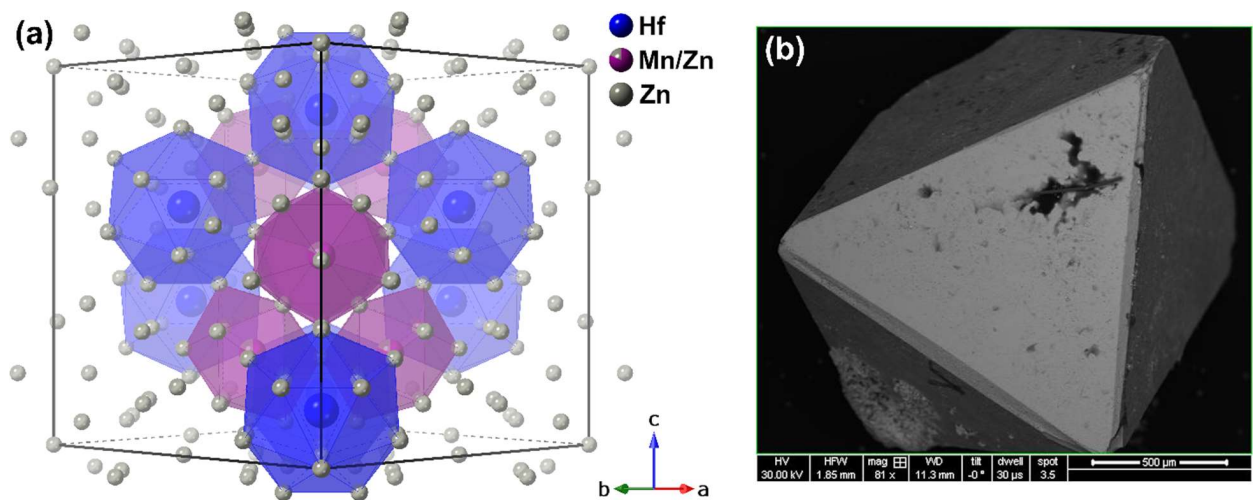

Figure 1. (a) Unit cell of $HfMn_2Zn_{20}$ viewed along [110]. The structure image was generated with CrystalMaker [49]. (b) SEM image of a crystal of $HfMn_2Zn_{20}$ in backscatter mode.

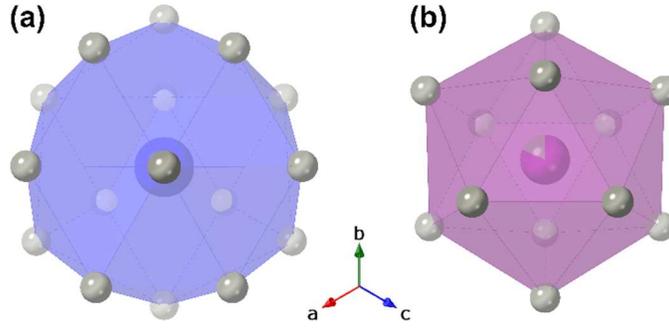

Figure 2: View along [111] of (a) the Frank-Kasper polyhedron with Hf in the center and (b) the icosahedron with Mn/Zn in the center. Colors are the same as in Figure 1.

The initial structural refinement based on the stoichiometric ratios resulted in good residuals ($R_1$ = 0.0273 and $wR_2$ = 0.0723), however, the elemental composition as obtained via EDS deviated enough from the stoichiometric 1:2:20 of $HfMn_2Zn_{20}$ to raise concerns, despite the fairly large error margin of EDS. Furthermore, Svanidze et al. [39] reported off-stoichiometry in the Zr-analog, $ZrMn_{1.78}Zn_{20.22}$, which exhibited Mn/Zn site mixing on the 16d site (the Mn site). We performed EDS measurements on several spots and areas on multiple crystals, and the average results are summarized in Table 1 together with calculated standard error of the mean. As can be seen, the Hf average value is within the ideal value. Mn, on the other hand, is lacking, and Zn is in excess. The EDS thus indicates a composition of $Hf_{1.03\pm0.05}Mn_{1.75\pm0.03}Zn_{20.21\pm0.04}$.

Table 1: Averages of the EDS results obtained on multiple crystals together with ratios based on the ideal 1:2:20 composition.

| Elements | Atomic percentage | Ideal ratios |
| --- | --- | --- |
| Hf | 4.5 ± 0.2 | 4.349 |
| Mn | 7.6 ± 0.1 | 8.696 |
| Zn | 87.9 ± 0.2 | 86.96 |
| | | |
| Composition | $Hf_{1.03\pm0.05}Mn_{1.75\pm0.03}Zn_{20.21\pm0.04}$ | $HfMn_2Zn_{20}$ |

Further refinement on the X-ray diffraction data revealed indeed a Mn/Zn mixing on the 16d site. Unlike the Zr-analog, however, we also observe an underoccupied 8a site (the Hf site). It is not uncommon to observe Zr mixing in Hf sites since there is typically a nominal amount of Zr in Hf chemicals. EDS, however, revealed no Zr. It is not clear why the Hf site is underoccupied as there is no clear reason for Zn to site-mix with Hf. Nevertheless, the residuals improved remarkably and all the crystal data and crystallographic parameters resulting from the single-crystal X-ray diffraction are summarized in Table 2. The final composition refined to $Hf_{0.93}Mn_{1.63}Zn_{20.37}$ ($Hf_{1-\delta}Mn_{2-}$ $_x$Zn$_{20+x}$, δ = 0.07, x = 0.37) which is fairly close to the composition found via EDS. For clarity, however, this compound will still be referred to as HfMn$_2$Zn$_{20}$. Table 3 lists the atomic coordinates.

Table 2: Single crystal X-ray diffraction data and parameters for HfMn$_2$Zn$_{20}$, collected at room temperature. The values in square brackets are powder X-ray diffraction data from the Rietveld refinement of crushed single crystals.

| Parameters | HfMn$_2$Zn$_{20}$ |
|---|---|
| Actual stoichiometry | Hf$_{0.93}$Mn$_{1.63}$Zn$_{20.37}$ (Hf$_{1-δ}$Mn$_{2-x}$Zn$_{20+x}$, δ=0.07, x=0.37) |
| Molecular weight (g/mol) | 1587.61 |
| Space group | $Fd\bar{3}m$ (#227) |
| $a$ (Å) | 14.0543(2) [14.0496] |
| $V$ (Å$^3$) | 2776.03(9) [2773.26] |
| Z | 8 |
| $ρ_{calc}$ (g/cm$^3$) | 7.597 [7.664] |
| Absorption coefficient $μ$ (mm$^{-1}$) | 42.744 |
| Absorption corrections $T_{min}$, $T_{max}$ | 0.22, 0.30 |
| Crystal size (mm$^3$) | 0.028 x 0.041 x 0.050 |
| Data collection range (°) | 2.510 < $θ$ < 44.778 |
| $h$ range | $-19 ≤ h ≤ 27$ |
| $k$ range | $-27 ≤ k ≤ 16$ |
| $l$ range | $-27 ≤ l ≤ 19$ |
| Reflections collected | 5375 |
| Independent reflections | 587 |
| Parameters refined | 18 |
| Restraints | 6 |
| Residual electron density $Δρ_{min}$, $Δρ_{max}$ (e/Å$^3$) | -6.36, 3.01 |
| $R_{int}$ | 0.023 |
| $R_1(F)$ for all data[a] | 0.0176 [0.0277] |
| $wR_2(F_o^2)$[b] | 0.0305 [0.0404] |
| Goodness-of-fit on $F^2$ | 1.0000 [1.67] |
| CSD # | 2212294 |

[a] $R_1 = Σ||F_o|-|F_c||/Σ|F_o|$ where $|F_o|$ refers to observed structure amplitude and $|F_c|$ to calculated structure amplitude.

[b] $wR_2 = [Σw(F_o^2 - F_c^2)^2 / Σw(F_o^2)^2]^{1/2}$, $w = 1/[σ^2(F_o^2) + (A·P)^2 + B·P]$, $P = [2F_c^2 + Max(F_o^2,0)]/3$ where $A = 0.01$ and $B = 22.81$.

Table 3. Atomic coordinates, and equivalent displacement parameters of HfMn$_2$Zn$_{20}$.

| Atom | Site | SOF | x | y | z | $U_{eq}$ (Å$^2$) |
| --- | --- | --- | --- | --- | --- | --- |
| Hf | 8a | 0.932(2) | 1/8 | 1/8 | 1/8 | 0.0063(5) |
| Mn11 | 16d | 0.816(9) | 1/2 | 1/2 | 1/2 | 0.0056(4) |
| Zn12 |  | 0.184(9) |  |  |  |  |
| Zn1 | 96g | 1 | 0.06010(2) | 0.06010(2) | 0.32312(2) | 0.0132(9) |
| Zn2 | 48f | 1 | 0.48833(2) | 1/8 | 1/8 | 0.0090(1) |
| Zn3 | 16c | 1 | 0 | 0 | 0 | 0.0148(1) |

Figure 3 displays the powder X-ray diffraction pattern obtained after grounding several clean crystals. The red line is a Rietveld refinement of the experimental data and guided by the lattice parameters obtained from the single-crystal X-ray diffraction refinement. All the observed peaks match well with the structure obtained from single-crystal X-ray diffraction refinement and only one spurious peak, belonging to residual zinc, was observed, indicating a single-phase compound. In other words, there were no magnetic impurity compounds present: if there were any such present, their amount was too small to be observed in the powder x-ray diffraction.

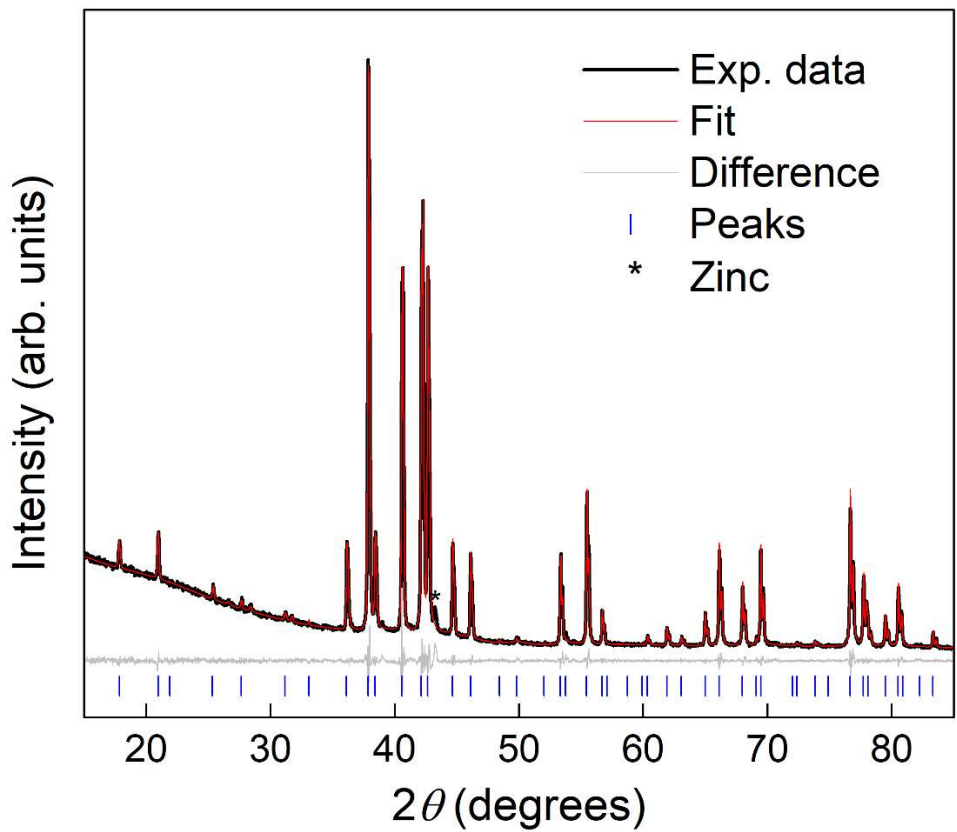

Figure 3: Powder X-ray diffraction pattern of ground HfMn$_2$Zn$_{20}$ crystals. The fit in red matches well with the pattern which indicates no major

impurities. Only a small additional peak indicating residual zinc was observed.

Gross *et al.* [5] compared the structural parameters of several HfM$_2$Zn$_{20}$ compounds including *M* = Fe, Co, and Ni where they showed a lattice constant and unit cell volume reduction consistent with the metallic radius trend of the transition metals [50]. We extend the trend with the addition of HfMn$_2$Zn$_{20}$. Figure 4 displays the trend in lattice parameter and unit cell volume of Hf*M*$_2$Zn$_{20}$ for *M* = Mn, Fe, Co, and Ni. While the metallic radius increases linearly from 1.24 Å for Ni to 1.27 Å for Mn [50], the lattice parameter and the unit cell volume are larger for our HfMn$_2$Zn$_{20}$ compound than expected from the linear trend. The reason for this deviation could be due to the Mn/Zn site mixing: the metallic radius of Zn is 1.34 Å, approximately 8.1% larger than the metallic radius of Mn, so that the inclusion of Zn on the Mn-site enlarges the overall unit cell of HfMn$_2$Zn$_{20}$ more than expected.

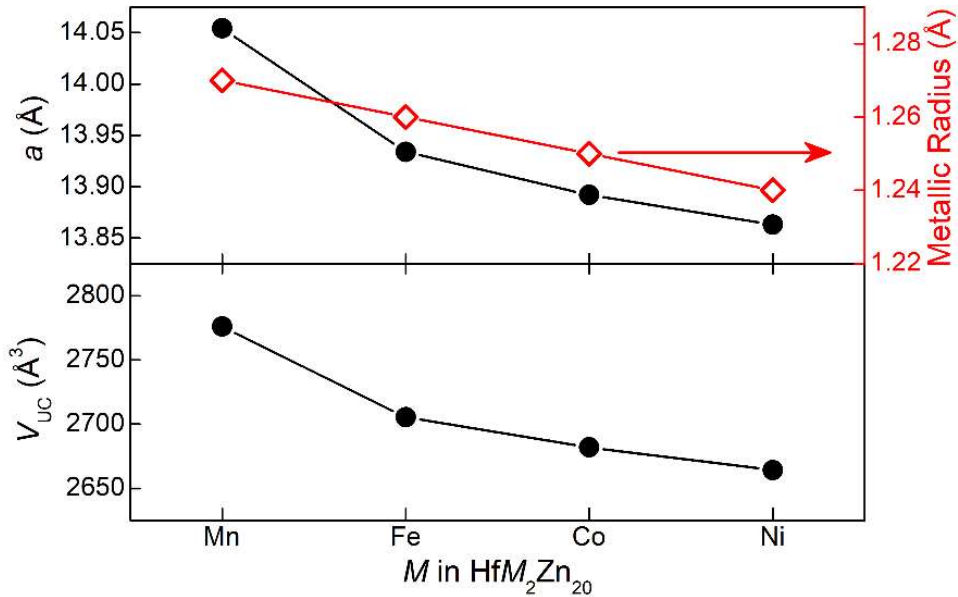

Figure 4: Lattice constant (upper panel) and unit cell volume (lower panel) comparison for Hf*M*$_2$Zn$_{20}$ (red circles) across *M* = Mn, Fe, Co, and Ni. The metallic radii are shown on the right axis of the upper panel (open, red diamonds) [50]. Data for the *M* = Fe, Co, and Ni compounds are taken from Ref. [5]. All error bars are within symbols.

In fact, looking at the volumes of the two polyhedra, we can observe the deviation of the Mn/Zn-containing icosahedron from the linear trend (upper panel in Figure 5). The Frank-Kasper polyhedron containing Hf, on the other hand, shrinks when *M* goes from Fe to Mn (lower panel in Figure 5). One reason could be the underoccupied Hf-site, although it is more likely that the overall structure is the real reason: a Frank-Kasper polyhedron is corner-shared with 12 $M$-containing icosahedra and only four other Frank-Kasper polyhedra. The Frank-Kasper polyhedron is therefore affected strongly by the relatively larger expansion of the surrounding icosahedra; hence, it is compressed.

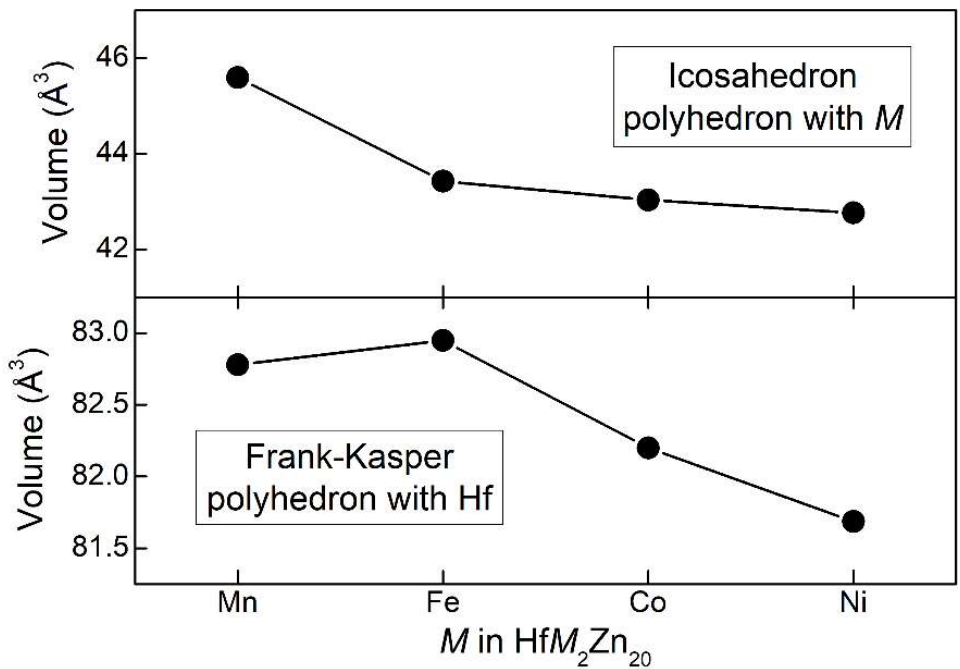

Figure 5: Volume of the icosahedron that contains the $M$ atom (upper panel) and volume of the Frank-Kasper polyhedron that contains Hf (lower panel) across Hf$M_2$Zn$_{20}$ with $M$ = Mn, Fe, Co, and Ni. Data for the $M$ = Fe, Co, and Ni compounds are taken from Ref. [5]. All error bars are within symbols.

*Magnetic properties*

Figure 6 displays the zero field-cooled magnetic susceptibility as a function of temperature at an applied field of 0.1 T. This field was chosen since low magnetization was expected due to the nature of the structure: Mn ions completely surrounded by non-magnetic Zn ions and with a large Mn–Mn distance of 4.969 Å. The derivative (solid line) indicates two transitions: one at approximately 22.3 K and one at approximately 8.7 K. A similar low-temperature transition was observed in the analogous ZrMn$_2$Zn$_{20}$ [39] at approximately 10 K which was attributed to ferromagnetic impurities. The transition appearing at 8.7 K in the derivative curve is likely due to the onset of saturation of the susceptibility, which is onsetting at approximately 16 K. Saturation of susceptibility is expected to take place in ferromagnets [51]. The saturation, however, is interrupted by a small upturn in susceptibility at approximately 3 K. This upturn is hard to explain: while a majority of the literature with magnetic measurements tends to explain such small upturns with the presence of small ferromagnetic impurities, our powder x-ray diffraction (Figure 3), indicated no presence of secondary phases. If any such impurities resulted from the synthesis, they must be incredibly small to not appear in the powder x-ray diffraction, and therefore contribute very little to the overall magnetization of this sample. Another reason for this upturn could be due to the appearance of a low-field saturation of the magnetization, discussed below. The magnetization behavior is clearly non-Curie-Weiss as can be seen by the inverse susceptibility as a function of temperature (inset to Figure 6). The lack of Curie-Weiss behavior together with two transitions were also observed in the ZrMn$_2$Zn$_{20}$ analog [39] where it was also noted that these behaviors were reminiscent of other compounds containing Mn [52]. The susceptibility, in fact, seems to follow the modified Curie-Weiss behavior [53]

$$\chi = \frac{C}{T - \theta} + \chi_0 \qquad (1)$$

in the region above the transition temperature. Here, $\theta$ is a Curie-Weiss temperature, $\chi_0$ is a temperature-independent term, and $C$ is the Curie constant and is related to the effective moment $\mu_{eff}$ via

$$C = \frac{N_A \mu_B^2 \mu_{eff}^2}{3k_B}. \qquad (2)$$

The susceptibility data above the transition temperature fits Eqs. (1) and (2) moderately well with the fit displayed in the large inset to Figure 6 as a green, solid line. The fit yields the effective moment $\mu_{eff} \approx 0.222(2)$ $\mu_B$/f.u., the Curie-Weiss temperature $\theta \approx 21.8(2)$ K, and $\chi_0 \approx 3.73(8)$ $\mu_B$/f.u., although the error margins are likely larger due to uncertainty in the sample mass. Nevertheless, the positive Curie-Weiss temperature indicates that the transition is possibly of a ferromagnetic nature with a $T_C \approx 22.3$ K as obtained from the derivative. Since the magnetic ion is manganese, the effective moment can be written $\mu_{eff} \approx 0.111(2)$ $\mu_B$/Mn if assuming stoichiometric HfMn$_2$Zn$_{20}$ or $\mu_{eff} \approx 0.137(2)$ $\mu_B$/Mn if assuming Hf$_{0.93}$Mn$_{1.63}$Zn$_{20.37}$. In any case, we note that the effective moment is quite small for Mn.

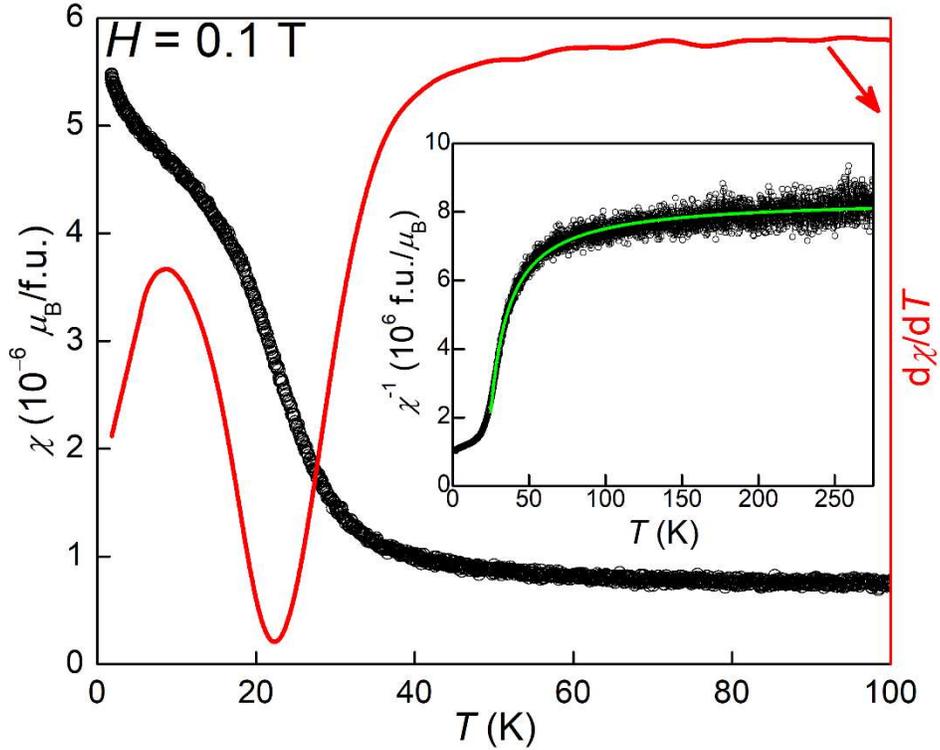

Figure 6: Zero field-cooled magnetic susceptibility as a function of temperature from 1.8 K to 300 K (although only up to 100 K is shown), obtained at an applied field of 0.1 T. The red solid line (right axis) is the derivative of the susceptibility. Inset: inverse susceptibility as a function of temperature. The green solid line is a fit to the modified Curie-Weiss law.

Figure 7 displays the magnetization as a function of applied field at 1.8 K and at 300 K. No high-field saturation is observed even at the lowest temperature (1.8 K) and highest applied field (6 T); the moment at 1.8 K and 6 T is quite small, approximately $9.9 \cdot 10^{-3}$ $\mu_B$/f.u. In addition, hysteresis in the 1.8 K isotherm is very small, essentially negligible (upper inset to Figure 7). All of these features – no saturation, small magnetic moment, and lack of hysteresis – have been taken as evidence for itinerant magnetism in the analogous $ZrMn_2Zn_{20}$ compound [39,42]. $HfMn_2Zn_{20}$ is, therefore, also a likely candidate for itinerant ferromagnetism.

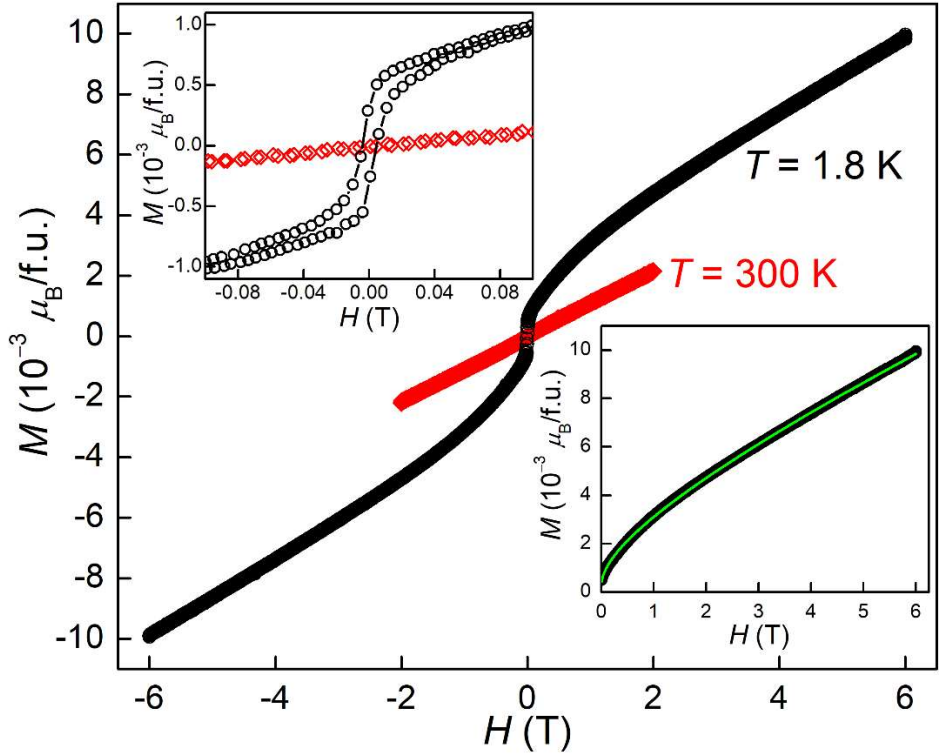

Figure 7: Magnetization as a function of applied field, obtained at 1.8 K (black circles) and 300 K (red diamonds). Upper inset: Close-up around zero field showing the low-field saturation and the near-negligible hysteresis at 1.8 K. Lower inset: Magnetization curve above zero applied field. The solid green line is a fit to Eq. 3 (see main text).

The isotherm at 1.8 K experiences a small but steep rise (upper inset to Figure 7) before a so-called low-field saturation (the bend in the isotherm) sets in at approximately 0.015 T. A spontaneous magnetization at 1.8 K, i.e., $M_{SP} = M(H = 0)$, can be obtained by fitting $M(H)$ above the low-field saturation to the expression [54,55]

$$M(H) = M_{SP} + a\sqrt{H} + \chi_{hf}H, \qquad (3)$$

where the middle term is related to the field-suppression of spin waves and $\chi_{hf}$ is the high-field susceptibility. The fit is displayed in the lower inset to Figure 7 as a green, solid line with parameters $M_{SP} \approx 2.84(5)\cdot 10^{-4}$ $\mu_B$/f.u., $a \approx 2.08(7)\cdot 10^{-3}$ $\mu_B$/f.u.·T$^{1/2}$, and $\chi_{hf} \approx 7.45(3)\cdot 10^{-4}$ $\mu_B$/f.u.·T. In these units, $M_{SP}$ is the spontaneous magnetic moment: $\mu_{SP} \approx 2.84(5)\cdot 10^{-4}$ $\mu_B$/f.u. This spontaneous magnetic moment is very small, as can also be seen in the upper inset to Figure 7. Extending the magnetization just above the initial bend linearly to zero field yields low values.

The nature of the magnetism can be probed further with the so-called Rhodes-Wohlfarth ratio $\mu_C/\mu_S$ where $\mu_C$ is the number of magnetic carriers in the paramagnetic state (i.e., at high temperatures) and $\mu_S$ is the number of magnetic carriers in the ordered state (i.e., at low temperatures) [42,56]. The Rhodes-Wohlfarth ratio shows how the number of magnetic carriers changes from high to low temperature. For local magnetism, $\mu_C/\mu_S = 1$, while for itinerant magnetism, $\mu_C/\mu_S \gg 1$. The $\mu_C$ is related to the paramagnetic moment and therefore the effective magnetic moment that was obtained in Eqns. (1) and (2). The relation can be expressed as [42,56,57]

$$\mu_{\text{eff}}^2 = \mu_C(\mu_C + 2) \qquad (4)$$

and yields $\mu_C \approx 0.0244$ $\mu_B$/f.u. The $\mu_S$ is related to the saturated magnetic moment via

$$\mu_{\text{sat}} = 2\mu_S \qquad (5)$$

and can be obtained at low temperatures and high fields. This is not the spontaneous magnetic moment that was obtained via Eqn. (3) since that value is related to the steep rise and low-field saturation. We, as stated above, do not observe an actual saturation of the magnetization (see Figure 7), a situation similar to the itinerant compound ZrMn$_2$Zn$_{20}$ [39]. Instead, we estimate by using the magnetization value at the highest field and lowest temperature we noted earlier: 0.03496 emu/g, corresponding to a "saturation" magnetic moment of 0.00994 $\mu_B$/f.u. This then yields $\mu_S \approx 0.00497$ $\mu_B$/f.u. With these values, we obtain the Rhodes-Wohlfarth ratio $\mu_C/\mu_S \approx 4.91$, which is reasonably large and clearly shows that the magnetism in HfMn$_2$Zn$_{20}$ is of itinerant nature. This value is added to the Rhodes-Wohlfarth plot of $\mu_C/\mu_S$ vs. $T_C$ [57] depicted in Figure 8(a), comparing several itinerant magnetism compounds.

The itinerant magnetism of this sample can be probed further with the spin fluctuation theories developed by Takahashi and Moriya [57-64] which built on the earlier Stoner criterion [65]. These theories have been explored in detail for several itinerant compounds [55,57,62-64,66-72]. Starting from minimization of the Landau expansion of the free energy, a relation is found between the magnetization and the applied field,

$$H = \frac{F_1}{N_A^3(2\mu_B)^4} M(M^2 - M_{SP}^2), \qquad (6)$$

where $M_{SP}$ is the spontaneous magnetization and $F_1$ is a mode-mode coupling term and expressed in terms of spin fluctuation parameters $T_0$ and $T_A$ [57,62,64]. Here, $T_0$ represents the energy width of the dynamical spin fluctuation spectrum and $T_A$ the dispersion of the static magnetic susceptibility in wave vector space. Importantly, $F_1$ can be evaluated experimentally via

$$F_1 = \frac{N_A^3 (2\mu_B)^4}{k_B \zeta}, \qquad (7)$$

where $\zeta \approx 6.83 \cdot 10^{-9}$ ($\mu_B$/f.u.)$^3$/T is the slope of the Arrott plot $M^2$ vs. $H/M$ [73] (not shown). The spin fluctuation parameters can then be expressed in terms of $T_C$, $M_{SP}$, and $F_1$ as [57,62,64]

$$\left(\frac{T_C}{T_0}\right)^{5/6} = \frac{M_{SP}^2}{40 C_{4/3}} \left(\frac{15 F_1}{T_C}\right)^{1/2} \qquad (8)$$

$$\left(\frac{T_C}{T_A}\right)^{5/3} = \frac{M_{SP}^2}{20 C_{4/3}} \left(\frac{4 T_C}{15 F_1}\right)^{1/3} \qquad (9)$$

where $C_{4/3} \approx 1.006089$ is a constant. With these expressions, Takahashi [57] showed that the degree of localization or itinerancy can be characterized by the ratio $T_C/T_0$: if the ratio is equal to unity, the compound has local magnetism while if $T_C/T_0 \ll 1$, the compound has itinerant magnetism. Using Eqns. (7) and (8), together with the $T_C$ found from Figure (6) and the $M_{SP}$ obtained via Eqn. (3), we obtain the ratio $T_C/T_0 \approx 0.0043$, thus, again proving that HfMn$_2$Zn$_{20}$ experiences itinerant magnetism. Furthermore, a generalized Rhodes-Wohlfarth plot – a Deguchi-Takahashi plot – was constructed out of experimental data showing the relation [57]

$$\frac{\mu_{\text{eff}}}{\mu_S} \approx 1.4 \left(\frac{T_C}{T_0}\right)^{-2/3}. \qquad (10)$$

Using the calculated ratio of $T_C/T_0$, we obtain $\mu_{\text{eff}}/\mu_S \approx 53$ and using the previously calculated $\mu_{\text{eff}}$ and $\mu_S$, we obtain $\mu_{\text{eff}}/\mu_S \approx 45$, in fairly good agreement with each other. The Deguchi-Takahashi plot [57,66] is shown in Figure 8(b) with our value added.

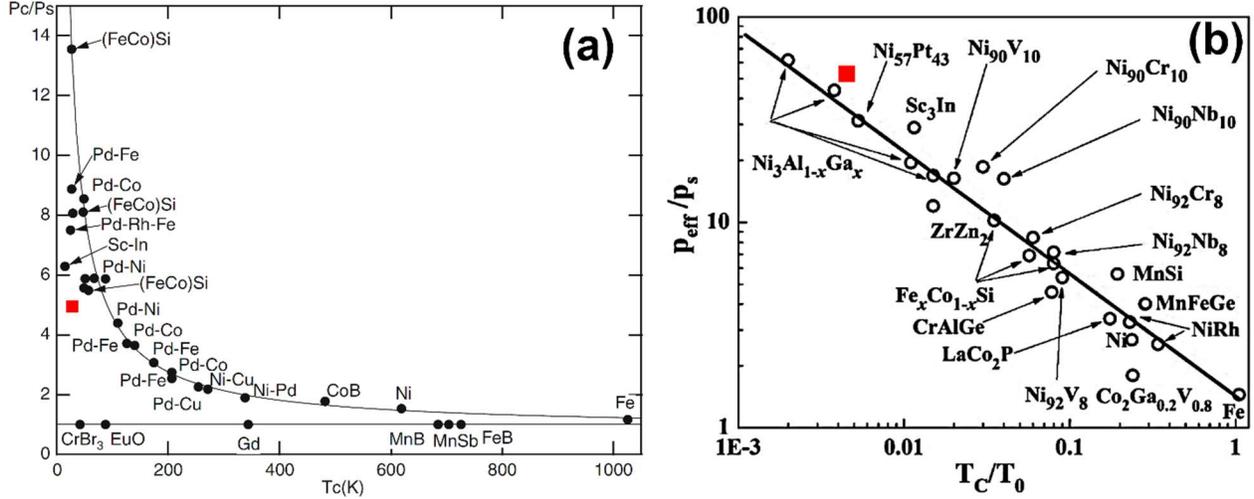

Figure 8: (a) Rhodes-Wohlfarth plot of $\mu_C/\mu_S$ vs. $T_C$ for various compounds with our compound as a red square. The plot was obtained from Ref. [57]. (b) Deguchi-Takahashi plot of $\mu_{eff}/\mu_S$ vs. $T_C/T_0$ for various compounds with our compound as a red square. The plot was obtained from Ref. [66]. Note that the authors in both references use the letter $p$ to denote magnetic moment while we employ the more commonly used $\mu$.

**Conclusion**

In conclusion, we have discovered a new member of the $AB_2C_{20}$ family: $HfMn_2Zn_{20}$. Due to Mn/Zn mixing on the Mn-site and an underoccupied Hf-site, the final stoichiometry is $Hf_{0.93}Mn_{1.63}Zn_{20.37}$ ($Hf_{1-\delta}Mn_{2-x}Zn_{20+x}$, $\delta = 0.07$, $x = 0.37$). This Mn-site mixing was also observed in the Zr-analog. The structure of the new compound follows expected lattice size trends when compared to existing isostructures with Hf and first-row transition metals. The magnetic susceptibility follows the modified Curie-Weiss law with a transition temperature around 22 K. The magnetization as a function of applied field shows no saturation, a small magnetic moment, and near negligible hysteresis, all signs of itinerant magnetism. The Rhodes-Wohlfarth ratio $\mu_C/\mu_S$ is approximately 4.91 and the spin fluctuation parameters ratio $T_C/T_0$ approximately 0.0043, both of these confirming the itinerant nature of the magnetism in $HfMn_2Zn_{20}$.

**Acknowledgement**

The authors acknowledge the NSF MRI program that funded the purchase of the Synergy-S X-ray diffractometer via award CHE-2117129. This work made use of a Quantum Design MPMS-3 supported by NSF (DMR-1920086) and the Cornell Center for Materials Research Shared Facilities which are supported through the NSF MRSEC program (DMR-1719875). T.B. also acknowledges financial support provided by Missouri State University's new faculty startup fund.